
------------------------------ Start of body part 2

%
%
%
%
%
%
%

\magnification=1200
for double spacing

\font\gross=cmbx10 scaled \magstep2
\font\mittel=cmbx10 scaled\magstep1
\font\pl=cmssq8 scaled \magstep1
\font\sc=cmcsc10

\def\RR{\rm I\!R}

\def\h#1{{\cal #1}}
\def\a{\alpha}
\def\b{\beta}
\def\g{\gamma}
\def\d{\delta}

\def\m{\mu}
\def\n{\nu}

\def\s{\sigma}
\def\om{\omega}
\def\na{\nabla}

\def\sq{\Square}
\def\square#1{\mathop{\mkern0.5\thinmuskip\vbox{\hrule
    \hbox{\vrule\hskip#1\vrule height#1 width 0pt\vrule}\hrule}
    \mkern0.5\thinmuskip}}
\def\Square{\mathchoice{\square{6pt}}{\square{5pt}}
    {\square{4pt}}{\square{3pt}}}


{\nopagenumbers
\null
\vskip-1.5cm
\hskip5cm{ \hrulefill }
\vskip-.55cm
\hskip5cm{ \hrulefill }
\vskip0.5mm
\rightline{ {\pl  University of Greifswald (May, 1994)}}
\vskip0.1mm
\hskip5cm{ \hrulefill }
\vskip-.55cm
\hskip5cm{ \hrulefill }
\bigskip
\rightline{ hep-th/9509079}
\bigskip
\rightline{ {\pl PHYSICS LETTERS B, {\bf 336} (1994) 171-177}}
\vfill

\centerline{\gross The heat kernel on symmetric spaces }
\medskip
\centerline{\gross via integrating over the group of isometries }
\bigskip
\bigskip
\centerline{{\mittel I. G. Avramidi}
\footnote{*}{ Alexander von Humboldt Fellow}
\footnote{$\S$}{On leave of absence from Research Institute for
Physics,
Rostov State University, Stachki 194, Rostov-on-Don 344104,
Russia}}

\centerline{\it Department of Mathematics, University of
Greifswald}
\centerline{\it Jahnstr. 15a, 17489 Greifswald, Germany}
\centerline{\sl E-mail: avramidi@math-inf.uni-greifswald.d400.de}

\bigskip
\smallskip
\vfill
\centerline{\sc Abstract}
\bigskip
{\narrower
{A new algebraic approach for calculating the heat kernel for the
Laplace operator on any Riemannian manifold with covariantly
constant curvature is proposed. It is shown that the heat kernel
operator can be obtained by an averaging over the Lie group of
isometries. The heat kernel diagonal is obtained in form of an
 integral over the isotropy subgroup.
\bigskip}}
\eject}


\leftline{\bf 1. Introduction}
\bigskip
The heat kernel $U(t)=\exp(t\sq)$ associated with the Laplace-Beltrami
 operator $\sq$ acting on a $d$-dimensional Riemannian manifold $M$
 without boundary with the metric $g_{\m\n}$ of Euclidean signature
 is of fundamental importance in mathematical physics [1-21]. In
 present paper we consider only the heat kernel diagonal
$$
U(t)\Big\vert_{diag} = \exp(-tH)g^{-1/2}\d(x,x')\Big\vert_{x=x'}
\eqno(1)
$$
and restrict ourselves to manifolds with covariantly constant
curvature
$$
\na_\m R_{\a\b\g\d} = 0. \eqno(2)
$$

We are {\it not} going to investigate in this paper the effects of
the {\it topology} but concentrate our attention on the {\it local}
effects. We calculate the heat kernel diagonal rather formal, having
in mind to get the {\it asymptotic expansion} at $t\to 0$ that does
not depend on the global structure of the manifold at all. This will
 then mean that we have obtained {\it all} the terms without
 derivatives of the curvature in the asymptotic expansion of the
 heat kernel.


The condition (2) determines the geometry of the symmetric spaces
[22]. The frame components of the curvature tensor (with respect
to some frame $e^a_\m$ that is parallel along geodesics) are constant
and can be presented in the form
$$
R_{abcd} = \b_{ik}E^i_{\ ab}E^k_{\ cd}, \eqno(3)
$$
where $E^i_{ab}$, $(i=1,\dots, p; p \le d(d-1)/2)$, is some set of
antisymmetric matrices. The matrix $\b_{ik}$ is known to be the
metric of the isotropy algebra $\h H$ with the generators
$D_i=\{D^a_{\ ib}\}$
$$
D^a_{\ ib}=-\b_{ik}E^k_{\ cb}g^{ca}= - D^a_{\ bi}, \eqno(4)
$$
$$
[D_i, D_k] = F^j_{\ ik} D_j, \eqno(5)
$$
$F^j_{\ ik}$ being the structure constants.

It is not difficult to show that the condition of integrability of the
equations (2) brings into existence a Lie algebra ${\cal G}$ of
dimension ${\rm dim}\,{\cal G}=D=p+d$, with structure constants
$C^A_{\ BC}=-C^A_{\ CB}$, $(A=1,\dots, D)$
$$
C^i_{\ ab}=E^i_{\ ab}, \quad C^a_{\ ib}=D^a_{\ ib},
\quad C^i_{\ kl}=F^i_{\ kl}, \eqno(6)
$$
$$
C^a_{\ bc}=C^i_{\ ka}=C^a_{\ ik}=0,
$$

Introducing a symmetric nondegenerate matrix
$$
\g_{AB} = \left(\matrix{ g_{ab} & 0             \cr
	       0                & \b_{ik}        \cr}\right),
	       \eqno(7)
$$
that plays the role of the metric on the algebra ${\cal G}$ one
can show that the adjoint and coadjoint representations of the
algebra ${\cal G}$ are equivalent, i.e.
$$
\g_{AB} C^B_{\ CD} + \g_{DB} C^B_{\ CA} = 0. \eqno(8)
$$

The generators of infinitesimal isometries $\xi_A=(P_a, L_i)$ of
symmetric spaces can be presented in the form [19,20]
$$
\eqalignno{
&P_{a} = P^\m_{\ a}\na_\m = -
\left(\sqrt K\cot\sqrt K\right)^b_{\ a}\h D_b, 		&(9)\cr
 &L_{i} = L^\m_{\ i}\na_\m = -D^b_{\ ia}\s^a\h D_b,
							&(10)\cr }
$$
where $\s^a(x,x')$ are the frame components of the tangent vector
to the geodesic between $x$ and $x'$, $\h D_a = {\partial /\partial
\s^a}$ and $K=\{K^a_{\ b}(x,x')\}$ is a matrix defined by
$$
K^a_{\ b}= R^a_{\ cbd}\s^c\s^d. \eqno(11)
$$
It is the algebra ${\cal G}$ that is generated by infinitesimal

isometries [22]
$$
[\xi_A,\xi_B]=C^C_{\ AB}\xi_C.    \eqno(12)
$$


\bigskip
\bigskip
\leftline{\bf 2. Heat kernel operator}
\bigskip

It is not difficult to show that the Laplacian in symmetric space
can be presented in terms of generators of isometries
$$
\sq = g^{\m\n}\na_\m\na_\n = \g^{AB}\xi_A\xi_B =g^{ab}P_a P_b
+ \b^{ik}L_i L_k,         \eqno(13)
$$
where $\g^{AB}=(\g_{AB})^{-1}$ and $\b^{ik}=(\b_{ik})^{-1}$.

Using this representation one can prove the following theorem [20].

\noindent
{\bf Theorem}.

\noindent
{\sl
For the {\it compact} Lie group $G$ (12) with positive definite metric
$\g_{AB}$ (7) and the structure constants $C^A_{\ BC}$ satisfying the
identity (8) it takes place the following identity for the Laplace
operator (13)
$$
\eqalignno{
\exp(t\sq) = &(4\pi t)^{-D/2} \int d k \g^{1/2}
	\det\left({\sinh(k^AC_A/2)\over k^AC_A/2}\right)^{1/2} &\cr
	& \times\exp\left\{ -{1\over 4t}k^A\g_{AB}k^B
	+ {1\over 6} R_G t\right\}\exp(k^A\xi_A) &(14)\cr}
$$
or, equivalently,
$$
\eqalignno{
\exp(t\sq) = &(4\pi t)^{-D/2} \int dq\,d\om \eta^{1/2}\b^{1/2}
\det\left({\sinh((q^a C_a + \om^i C_i)/2)\over (q^a C_a + \om^i C_i)/2}
\right)^{1/2} &\cr
& \times\exp\left\{ -{1\over 4 t}(q^a g_{ab}q^b + \om^i\b_{ik}\om^k)
+ \left({1\over 8} R + {1\over 6} R_H \right) t \right\}
\exp\left(q^a P_a + \om^i L_i\right) ,  &\cr
&                 &(15)\cr}
$$
where $\g=\det\g_{AB}$, $\b=\det \b_{ik}$, $\eta=\det g_{ab}$, $R_G$ is
the scalar curvature of the group $G$
$$
R_G= -{1\over 4}\g^{AB} C^C_{\ AD}C^D_{\ BC}
= {3\over 4} R + R_H ,                            \eqno(16)
$$
and $R_H$ is the scalar curvature of the isotropy subgroup $H$
$$
R_H = -{1\over 4} \b^{ik} F^m_{\ \ il}F^l_{\ km}, \eqno(17)
$$
with $\b^{ik}=(\b_{ik})^{-1}$,
the matrices $C_A = \{C^B_{\ AC}\} = ( C_a, C_i )$ are the generators
of the adjoint representation of the algebra defined by
$$
C_a = \left( \matrix{ 0          & D^b_{\ ai}   \cr
		      E^j_{\ ac} & 0            \cr}\right) ,\qquad
C_i = \left( \matrix{ D^b_{\ ia} & 0            \cr
		      0          & F^j_{\ ik}   \cr}\right) ,
						\eqno(18)
$$
and the integration is to be taken over the whole Euclidean
space $\RR^D$.}

\noindent
{\bf Proof}.
\vglue0pt
\noindent
First one can show that
$$
\sq\exp(k^A\xi_A) =  X_2 \exp(k^A\xi_A) \eqno(19)
$$
where $X_2=\g^{BC} X_B X_C$ and $ X_A =  X^M_{\ A} (k) {\partial/\partial
k^M} $ are the left-invariant vector fields on the group
$$
X^M_{\ A}(k) = \left({k^AC_A\over \exp(k^AC_A) -1}\right)^M_{\ A}.
							\eqno(20)
$$
Then, introducing the metric on the group manifold
$$
G_{MN} = \g_{AB} X^{-1 A}_{\ \ \ \ M} X^{-1 B}_{\ \ \ \ N} \eqno(21)
$$
and integrating by parts one has from (19)
$$
\eqalignno{
&\sq\int d k \g^{1/2}\Phi (t|k) \exp(k^A\xi_A) &\cr
&=\int d k \g^{1/2}\exp(k^A\xi_A)
\left(G^{1/2}X_2 G^{-1/2}\Phi(t|k) \right) . &(22)\cr}
$$
where $G=\det G_{MN}$ and
$$
\eqalignno{
\Phi(t|k) = &(4\pi t)^{-D/2}
		\det\left({\sinh(k^AC_A/2)\over k^AC_A/2}\right)^{1/2}
\exp\left\{ -{1\over 4t}k^A\g_{AB}k^B
	+ {1\over 6} R_G t\right\}.        &(23)\cr}
$$
Further, using the equations
$$
X_2 G^{-1/4} = {1\over 6} R_G G^{-1/4}              \eqno(24)
$$
and
$$
k^A{\partial\over \partial k^A} G^{-1/4} = {1\over 2} (D - X^A_{\ A})
G^{-1/4},                                               \eqno(25)
$$
that hold on the group manifold [21], one can show that $\Phi(t|k)$
satisfies the equation
$$
\partial_t \Phi = G^{1/2}X_2 G^{-1/2}\Phi           \eqno(26)
$$
and initial condition
$$
\Phi(t|k)\Big\vert_{t=0} = \g^{-1/2}\d(k).
\eqno(27)
$$
Therefrom it follows
$$
\left(\partial_t-\sq\right)
\int d k \g^{1/2}\Phi (t|k) \exp(k^A\xi_A)=0, \eqno(28)
$$
and hence
$$
\int d k \g^{1/2}\Phi (t|k) \exp(k^A\xi_A)=\exp(t\sq), \eqno(29)
$$
that proves the theorem.

\bigskip
\bigskip
\leftline{\bf 3. Heat kernel diagonal}
\bigskip

To get the heat kernel in coordinate representation one has to act
with the heat kernel operator $\exp(t\sq)$ (15) on the $\d$-function
(1). One can show that [20]
$$
\exp\left(q^a P_a + \om^i L_i\right)g^{-1/2}\d(x,x')\Big\vert_{x=x'}
=\eta^{-1/2}\d(\s^a_0(s,q,\om))\Big\vert_{s=1} \eqno(30)
$$
where $\s^a_0(s,q,\om)$ is to be determined from the equation of
characteristics
$$
{d \s_0^a\over ds} = - \left(\sqrt {K(\s_0)}\cot \sqrt
 {K(\s_0)}\right)^a_{\ b}q^b - \om^iD^a_{\ ib}\s^b_0 . \eqno(31)
$$
with initial condition
$$
\s^a_0\big\vert_{s=0}=0 \eqno(32)
$$
Solving the eq. (31) for small $\s^a_0$ we have
$$
\s_0^a(s,q,\om)\Big\vert_{s=1} = \left({\exp(-s \om^iD_i)-
1\over \om^iD_i}\right)^a_{\ b}q^b + O(q^2) \eqno(33)
$$
and, therefore,
$$
\d(\s_0^a(s,q,\om))\Big\vert_{s=1}=\det\left({\sinh(\om^iD_i/2)\over
\om^iD_i/2}\right)^{-1}\d(q).  \eqno(34)
$$

Now substituting (30) and (34) in (15) we can easily integrate over
$q$ to get finally the heat kernel diagonal
$$
\eqalignno{
U(t)\Big\vert_{diag} = &(4\pi t)^{-D/2}\int d\om \b^{1/2}
\det\left({\sinh(\om^iF_i/2)\over \om^iF_i/2}\right)^{1/2}
\det\left({\sinh(\om^iD_i/2)\over \om^iD_i/2}\right)^{-1/2} &\cr
&\times\exp\left\{ - {1\over 4 t}\om^i\b_{ik}\om^k
+ \left({1\over 8} R
+ {1\over 6} R_H \right) t \right\},                     &(35)\cr}
$$
where the matrices $F_i=\{F^k_{\ ij}\}$ are the generators of the
isotropy algebra in the adjoint representation.

One can present this result also in an alternative nontrivial rather
{\it formal} way. Substituting the equation
$$
(4\pi t)^{-p/2}\b^{1/2}\exp\left(-{1\over 4t} \om^i\b_{ik}\om^k\right) =
(2\pi)^{-p}\int dp \exp \left(ip_k\om^k -tp_k\b^{kn}p_n\right) \eqno(36)
$$
into the integral (35), integrating over $\om$ and changing the
integration variables $p_k \to i t^{-1/2} p_k $ we get finally an
expression without any integration
$$
\eqalignno{
U(t)\Big\vert_{diag}& = (4\pi t)^{-d/2}\exp\left(t\left({1\over 8} R
+ {1\over 6} R_H \right)\right) &\cr
&\times\det\left({\sinh(\sqrt t \partial^kF_k/2)\over \sqrt t
\partial^kF_k/2}\right)^{1/2}
\det\left({\sinh(\sqrt t \partial^kD_k/2)\over \sqrt t
\partial^kD_k/2}\right)^{-1/2}
\exp\left(p_n\b^{nk}p_k\right)\Bigg\vert_{p=0}. &\cr
&       &(37)\cr}
$$
where $\partial^k=\partial/\partial p_k$.

This formal solution should be understood as a power series in the
derivatives $\partial^i$ that is well defined and determines the heat
 kernel asymptotic expansion at $t\to 0$.

\bigskip
\bigskip
\leftline{\bf 4. Concluding remarks}
\bigskip

In present paper we proposed a new purely algebraic approach for
calculating the heat kernel diagonal in symmetric spaces that is
based essentially on the Lie group of isometries.

We proved a theorem that expresses the heat kernel operator $\exp(t\sq)$,
i.e. the exponential of the {\it second} order operator, in terms of the
isometries $\exp(k^A\xi_A)$, i.e. the exponential of {\it first} order
operator.

Let us mention, that our formulae for the heat kernel diagonal are
{\it exact} (up to possible nonanalytic topological contributions).
This gives a nontrivial example how the heat kernel can be constructed
using only the commutation relations of some differential operators,
namely the generators of infinitesimal isometries of the symmetric
space. These formulae can be used now to generate asymptotic expansion
of the heat kernel for {\it any} symmetric space, {\it any space with
covariantly constant curvature}, simply by expanding them in a power
series in $t$.

In present paper we considered for simplicity the case of symmetric
space of compact type, i.e. with positive curvature (positive definite
matrix $\b_{ik}$). There is a remarkable duality relation between the
compact and noncompact symmetric spaces [22, 20]
$R^*_{\a\b\g\d}=-R_{\a\b\g\d}$, $\b^*_{ik}=-\b_{ik}$, $E^{i*}_{\ ab}
=-E^i_{\ ab}$, $D^{a*}_{\ ib}=D^{a}_{\ ib}$, $F^{i*}_{\ jk}=F^{i*}_{\ jk}$.
Therefore, one can get the results for noncompact case by means
of analytic continuation. This means that our formulae (35) and (37)
should be valid in general case of arbitrary symmetric space that is
the product of compact, noncompact and the Euclidean ones. Moreover,
It should also be valid for the case of pseudo-Euclidean signature
of the metric $g_{\m\n}$. However, it is not perfectly clear {\it how}
to do the analytic continuation for obtaining physical results.

\bigskip
\bigskip
\leftline{\bf Acknowledgements }
\bigskip

I would like to thank G. A. Vilkovisky for many helpful discussions
and R. Schimming and J. Eichhorn for their hospitality at the
University of Greifswald. I am
also grateful to P. B. Gilkey, H. Osborn, S. Fulling, T. Osborn,
S. Odintsov and K.  Kirsten for correspondence.
This work was supported, in part, by a Soros Humanitarian Foundations
Grant
awarded by the American Physical Society and by an Award
  through the International Science Foundation's Emergency Grant
  competition.

\bigskip\bigskip
\leftline{\bf References}
\bigskip

\item{[1]}  B. S. De Witt,
		Dynamical theory of groups and fields (Gordon and
		Breach,
		New York, 1965);
		in: Relativity, groups and topology II, ed. by B. S.
		De Witt
		and R. Stora (North Holland, Amsterdam, 1984) p. 393
\item{[2]} G. A. Vilkovisky,
		in: Quantum theory of gravity, ed.  S. Christensen
		(Hilger,
		Bristol, 1983) p. 169
\item{[3]} A. O. Barvinsky and G. A. Vilkovisky,
		Phys. Rep. C 119 (1985) 1
\item{[4]} I. G. Avramidi,
		Nucl. Phys. B 355 (1991) 712
\item{[5]} P. B. Gilkey,
		Invariance theory, the heat equation and the  Atiyah
		- Singer
		index theorem (Publish or Perish, Wilmington, DE, USA,
		1984)
\item{[6]} J. Hadamard,
		Lectures on Cauchy's Problem, in: Linear Partial
		Differential
		Equations (Yale U. P., New Haven, 1923)
\item{} S. Minakshisundaram and A. Pleijel,
		Can. J. Math. 1 (1949) 242
\item{} R. T. Seeley,
		Proc. Symp. Pure Math. 10 (1967) 288
\item{} H. Widom,
		Bull. Sci. Math. 104 (1980) 19
\item{} R. Schimming,
		Beitr. Anal. 15 (1981) 77;
		Math. Nachr. 148 (1990) 145
\item{[7]} N. E. Hurt,
		Geometric quantization in action:  applications of
		harmonic
		analysis in quantum statistical mechanics and quantum
		field
		theory, (D. Reidel Publishing Company,  Dordrecht,
		Holland,
		1983)
\item{[8]} P. B. Gilkey,
		Functorality and heat equation asymptotics, in:
		Colloquia
		Mathematica Societatis Janos Bolyai, 56. Differential
		Geometry, (Eger (Hungary), 1989), (North-Holland,
		Amsterdam,
		1992), p.  285
\item{} Forty More Years of Ramifications: Spectral Asymptotics and
Its Applications, Ed. by S. A. Fulling and F. J. Narcowich, Discourses
in Mathematics and Its Applications, No 1, Department of Mathematics,
Texas A \& M University, College Station, Texas, 1991
\item{} R. Schimming,
		Calculation of the heat kernel coefficients, in: B.
		Riemann
		Memorial Volume, ed. T. M.  Rassias, (World
		Scientific,
		Singapore), to be published
\item{[9]} P. B. Gilkey,
		J. Diff. Geom. 10 (1975) 601
\item{[10]} I. G. Avramidi,
		Teor. Mat. Fiz. 79 (1989) 219;
		Phys. Lett. B 238 (1990) 92
\item{[11]} P. Amsterdamski, A. L. Berkin and D. J. O'Connor,
		Class. Quantum  Grav. 6 (1989) 1981
\item{[12]} T. P. Branson  and P. B. Gilkey,
		Comm. Part. Diff. Eq. 15 (1990) 245
\item{}     N. Blazic, N. Bokan and P. B. Gilkey, Indian J. Pure
Appl. Math.
		23 (1992) 103
\item{}     M. van den Berg and P. B. Gilkey, Heat content asymptotics
of a
		Riemannian manifold with boundary, University of
		Oregon
		preprint (1992)
\item{}     S. Desjardins and P. B. Gilkey, Heat content asymptotics
for
		operators of Laplace type with Neumann boundary
		conditions,
		University of Oregon preprint (1992)
\item{}     M. van den Berg, S. Desjardins and P. B. Gilkey,
Functorality and
		heat content asymptotics for operators of Laplace
		type,
		University of Oregon preprint (1992)
\item{[13]}         G. Cognola, L. Vanzo and S. Zerbini,
		Phys. Lett. B 241 (1990) 381
\item{}     D. M. Mc Avity and H. Osborn,
		Class. Quantum Grav. 8 (1991) 603;
		Class. Quantum Grav. 8 (1991) 1445;
		Nucl. Phys. B 394 (1993) 728
\item{}     A. Dettki and A. Wipf,
		Nucl. Phys. B 377 (1992) 252
\item{}     I. G. Avramidi,
		Yad. Fiz. 56 (1993) 245

\item{[14]} S. A. Fulling,
		SIAM J. Math. Anal. 13 (1982) 891;
		J. Symb. Comput. 9 (1990) 73
\item{}     S. A. Fulling and G. Kennedy,
		Trans. Am. Math. Soc. 310 (1988) 583
\item{}  V. P. Gusynin,
		Phys. Lett. B 255 (1989) 233

\item{[15]} I. G. Avramidi,
		Yad. Fiz. 49 (1989) 1185;
		Phys. Lett. B 236 (1990) 443
\item{[16]}     T. Branson, P. B. Gilkey and B. \O rsted,
		Proc. Amer. Math. Soc. 109 (1990) 437
\item{[17]} A. O. Barvinsky and G. A. Vilkovisky,
		Nucl. Phys. B 282 (1987) 163;
		Nucl. Phys. B 333 (1990) 471
\item{}     G. A. Vilkovisky,
		Heat kernel: recontre entre physiciens et
		mathematiciens,
		preprint CERN-TH.6392/92 (1992), in: Proc. of
		Strasbourg
		Meeting between physicists and  mathematicians
		(Publication de
		l' Institut de Recherche Math\'ematique  Avanc\'ee,
		Universit\'e  Louis
		Pasteur, R.C.P. 25, vol.43 (Strasbourg, 1992)), p. 203
\item{}     A. O. Barvinsky, Yu. V. Gusev, V. V. Zhytnikov and G. A.
		Vilkovisky, Covariant perturbation theory (IY), Report
		of the University of Manitoba (University of Manitoba,
		Winnipeg, 1993)
\item{[18]} I. G. Avramidi,
		Phys. Lett. B 305 (1993) 27
\item{[19]} I. G. Avramidi,
		Covariant methods for calculating the low-energy
		effective
		action in quantum field theory and quantum gravity,
		University of Greifswald (1994), gr-qc/9403036
\item{[20]} I. G. Avramidi,
		A new algebraic approach for calculating the heat
		kernel in
		quantum gravity,
		University of Greifswald, EMA-MAT-1994-5,
		hep-th/9406047
\item{[21]} J. S. Dowker,
		Ann. Phys. (USA) 62 (1971) 361;
		J. Phys. A 3 (1970) 451
\item{} A. Anderson and R. Camporesi,
		Commun. Math. Phys. 130 (1990) 61
\item{} R.  Camporesi,
		Phys. Rep. 196 (1990) 1
\item{[22]} H. S. Ruse, A. G. Walker, T. J. Willmore,
		Harmonic spaces, (Edizioni Cremonese, Roma (1961))
\item{} J. A. Wolf,
		Spaces of constant curvature
		(University of California, Berkeley, CA, 1972)
\item{}  B. F. Dubrovin, A. T. Fomenko and S. P. Novikov,
		The Modern  Geometry: Methods and  Applications
		(Springer, N.Y.
		1992)
\bye